\documentclass[a4paper,11pt]{article}
\pdfoutput=1 % if your are submitting a pdflatex (i.e. if you have
\usepackage{jinstpub} % for details on the use of the package, please
\usepackage{titlesec}                     % see the JINST-author-manual
\usepackage{parskip}
\usepackage{float}
\usepackage{bold-extra}
\usepackage{amsmath}
\usepackage{amssymb}
\usepackage{times, txfonts}
\titlespacing{\section}{0pt}{\parskip}{2pt}
\titlespacing{\subsection}{0pt}{\parskip}{0pt}
\titlespacing{\subsubsection}{0pt}{\parskip}{-\parskip}

\author[a,1]{M. A. Shah,\note{Corresponding author.}}
\author[a]{A. Ahmad,}
\author[a]{M. Ahmad,}
\author[a]{S. Muhammad,}
\author[a]{H. R. Hoorani,}
\author[a]{M. I. Asghar,}
\author[a]{I. M. Awan,}
\affiliation[a]{National Centre for Physics, Quaid-i-Azam University, Islamabad, Pakistan}

\author[b]{S. Aly,}
\author[b]{Y. Assran,}
\author[b]{A. Radi,}
\author[b]{A. Sayed,}
\affiliation[b]{Egyptian Network for High Energy Physics, Academy of Scientific Research and Technology, 101 Kasr El-Einy St. Cairo Egypt}

\author[c]{G. Singh,}
\affiliation[c]{Chulalongkorn University, Department of Physics, Faculty of Science, Payathai Road, Phatumwan, Bangkok, THAILAND}

\author[d]{M. Abbrescia,}
\author[d]{G. Iaselli,}
\author[d]{M. Maggi,}
\author[d]{G. Pugliese,}
\author[d]{P. Verwilligen,}
\affiliation[d]{INFN, Sezione di Bari, Via Orabona 4, IT-70126 Bari, Italy}

\author[e]{W. V. Doninck,}
\affiliation[e]{Vrije Universiteit Brussel, Boulevard de la Plaine 2, 1050 Ixelles, Belgium}

\author[f]{S. Colafranceschi,}
\author[f]{A. Sharma,}
\affiliation[f]{Physics Department CERN, CH-1211 Geneva 23, Switzerland}

\author[g]{L. Benussi,}
\author[g]{S. Bianco,}
\author[g]{D. Piccolo,}
\author[g]{F. Primavera,}
\affiliation[g]{INFN, Laboratori Nazionali di Frascati (LNF), Via Enrico Fermi 40, IT-00044 Frascati, Italy}

\author[h]{V. Bhatnagar,}
\author[h]{R. Kumari,}
\author[h]{A. Mehta,}
\author[h]{J. Singh,}
\affiliation[h]{Department of Physics, Panjab University, Chandigarh Mandir 160 014, India}

\author[i]{A. Cimmino,}
\author[i]{S. Crucy,}
\author[i]{A. Fagot,}
\author[i]{M. Gul,} 
\author[i]{A. A. O. Rios,}
\author[i]{M. Tytgat,}
\author[i]{N. Zaganidis,}
\affiliation[i]{Ghent University, Department of Physics and Astronomy, Proeftuinstraat 86, 9000 Ghent, Belgium}

\author[j]{S. W. Cho,}
\author[j]{S. Y. Choi,}
\author[j]{B. Hong,}
\author[j]{M. H. Kang,}
\author[j]{K. S. Lee,}
\author[j]{J. H. Lim,}
\author[j]{S. K. Park,}
\affiliation[j]{Korea University, Department of Physics, 145 Anam-ro, Seongbuk-gu,Seoul 02841, Republic of Korea}

\author[k]{M. S. Kim,}
\affiliation[k]{Kyungpook National University, 80 Daehak-ro, Buk-gu, Daegu 41566, Republic of Korea}

\author[l]{M. Goutzvitz,}
\author[l]{G. Grenier,}
\author[l]{F. Lagarde,}
\author[l]{F. Lagarde,}
\affiliation[l]{Universite de Lyon, Universite Claude Bernard Lyon 1, CNRS-IN2P3, Institut de Physique Nucleaire de Lyon, Villeurbanne, France}

\author[m]{C. U. Estrada,}
\author[m]{I. Pedraza,}
\author[m]{C. B. Severiano,}
\affiliation[m]{Benemerita Universidad Autonoma de Puebla, Puebla, Mexico}

\author[n]{S. Carrillo Moreno,}
\author[n]{F. Vazquez Valencia,}
\affiliation[n]{Universidad Iberoamericana, Mexico City, Mexico}

\author[o]{L. M. Pant,}
\affiliation[o]{Nuclear Physics Division Bhabha Atomic Research Centre Mumbai 400 085, INDIA}

\author[p]{S. Buontempo,}
\author[p]{N. Cavallo,}
\author[p]{M. Esposito,}
\author[p]{F. Fabozzi,}
\author[p]{G. Lanza,}
\author[p]{L. Lista,}
\author[p]{S. Meola,}
\author[p]{M. Merola,}
\author[p]{I. Orso,}
\author[p]{P. Paolucci,}
\author[p]{F. Thyssen,}
\affiliation[p]{INFN, Sezione di Napoli, Complesso Univ. Monte S. Angelo, Via Cintia, IT-80126 Napoli, Italy} 

\author[q]{A. Braghieri,}
\author[q]{A. Magnani,}
\author[q]{P. Montagna,}
\author[q]{C. Riccardi,}
\author[q]{P. Salvini,}
\author[q]{I. Vai,}
\author[q]{P. Vitulo,}
\affiliation[q]{INFN, Sezione di Pavia, Via Bassi 6, IT-Pavia, Italy} 

\author[r]{Y. Ban,}
\author[r]{S. J. Qian,}
\affiliation[r]{School of Physics, Peking University, Beijing 100871, China} 

\author[s]{M. Choi,}
\affiliation[s]{University of Seoul, 163 Seoulsiripdae-ro, Dongdaemun-gu, Seoul, Republic of Korea}

\author[t]{Y. Choi,}
\author[t]{J. Goh,}
\author[t]{D. Kim,}
\affiliation[t]{Sungkyunkwan University, 2066 Seobu-ro, Jangan-gu, Suwon-si, Gyeonggi-do, Republic of Korea}

\author[u]{A. Aleksandrov,}
\author[u]{R. Hadjiiska,}
\author[u]{P. Iaydjiev,}
\author[u]{M. Rodozov,}
\author[u]{S. Stoykova,}
\author[u]{G. Sultanov,}
\author[u]{M. Vutova,}
\affiliation[u]{Bulgarian Academy of Sciences, Inst. for Nucl. Res. and Nucl. Energy, Tzarigradsko shaussee Boulevard 72, BG-1784 Sofia, Bulgaria}

\author[v]{A. Dimitrov,}
\author[v]{L. Litov,}
\author[v]{B. Pavlov,}
\author[v]{P. Petkov,}
\affiliation[v]{Faculty of Physics, University of Sofia,5, James Bourchier Boulevard, BG-1164 Sofia, Bulgaria}

\author[w]{D. Lomidze,}
\author[w]{I. Bagaturia,}
\affiliation[w]{Tbilisi University, 1 Ilia Chavchavadze Ave, Tbilisi 0179, Georgia}

\author[x]{C. Avila,}
\author[x]{A. Cabrera,}
\author[x]{J. C. Sanabria,}
\affiliation[x]{Universidad de Los Andes, Apartado Aereo 4976, Carrera 1E, no. 18A 10, CO-Bogota, Colombia}

\author[y]{I. Crotty,}
\affiliation[y]{Dept. of Physics, Wisconsin University, Madison, WI 53706, United States}

\author[z]{and J. Vaitkus}
\affiliation[z]{Vilnius University, Vilnius, Lithuania}

\title{\boldmath Comparison of CMS Resistive Plate Chambers performance during LHC RUN-1 and RUN-2}

\emailAdd{mashah@cern.ch}

\abstract{The Resistive Plate Chambers detector system at the CMS experiment at the LHC provides robustness and redundancy to the muon trigger. A total of 1056 double-gap chambers cover the pseudo-rapidity region |$\eta$| $\le$ 1.6. The main detector parameters and environmental conditions are constantly and closely monitored to achieve operational stability and high quality data in the harsh conditions of the second run period of the LHC with center-of-mass energy ($\sqrt s$) = 13 TeV. First results of overall detector stability with 2015 data and comparisons with data from the LHC RUN-1 period at $\sqrt s$ = 8 TeV are presented.}

\keywords{Resistive-plate chambers}

\collaboration[c]{on behalf of the CMS Collaboration}
\proceeding{13$^{\text{th}}$ Workshop on Resistive Plate Chambers and Related Detectors\\
  February 22-26, 2016\\
  Ghent University, Belgium}

\begin{document}
\maketitle
\flushbottom
\section{Introduction}
\label{sec:intro}
\subsection{CMS Experiment}
The Large Hadron Collider (LHC)~\cite{a}, the most energetic particle accelerator ever built is a double ring structure that collides beams of protons at a center-of-mass energy of 13 TeV. Located at one of the 4 interaction points is the Compact Muon Solenoid (CMS) experiment~\cite{b}, where three types of gaseous detectors are used to identify and measure muons. In the barrel region (the pseudo-rapidity |$\eta$| < 1.2), Drift Tube (DT) chambers are used. In the endcaps, Cathode Strip Chambers (CSC) are deployed which cover the region up to |$\eta$| < 2.4. In addition to this, resistive plate chambers (RPC)~\cite{c} are installed in both barrel and end-cap regions. These RPCs are operated in the avalanche mode to ensure the expected time resolution of $\approx$ 2 ns at rates of the order of 10 kHz/cm$^2$. To monitor and optimise the performance of RPCs within CMS, data taken during 2015 at 13 TeV have been studied and compared with 2012 data at 8 TeV. 
\subsection{The Resistive Plate Chambers}
The CMS-RPC system is composed of double-gap chambers, each 2 mm gas gap formed by two parallel bakelite electrodes. Copper readout signal electrodes are placed in between the gas-gaps. In the barrel the muon system is made of four coaxial stations, interleaved with iron yokes. The endcap region consists of three iron disks interlayed with 4 RPC stations.
The geometry of the RPC strips is mainly driven by the need to have the trigger adjustable on different pT muons. In the barrel, the strip shapes are rectangular while in the endcaps they are trapezoidal. More details about the CMS muon system might be found in the Technical Design Report ~\cite{d}. The total number of readout channels of the RPC system is larger than 130,000. %, covering a total area of 2953m$^2$. 
\section{RPC Performance during RUN-1 and RUN-2}
\subsection{RPC Background}
Background radiation level in the CMS muon system is one of the important factors in the overall performance. 
\begin{figure}[H]
\centering
\hspace{-0.5cm}
\includegraphics[scale=0.6]{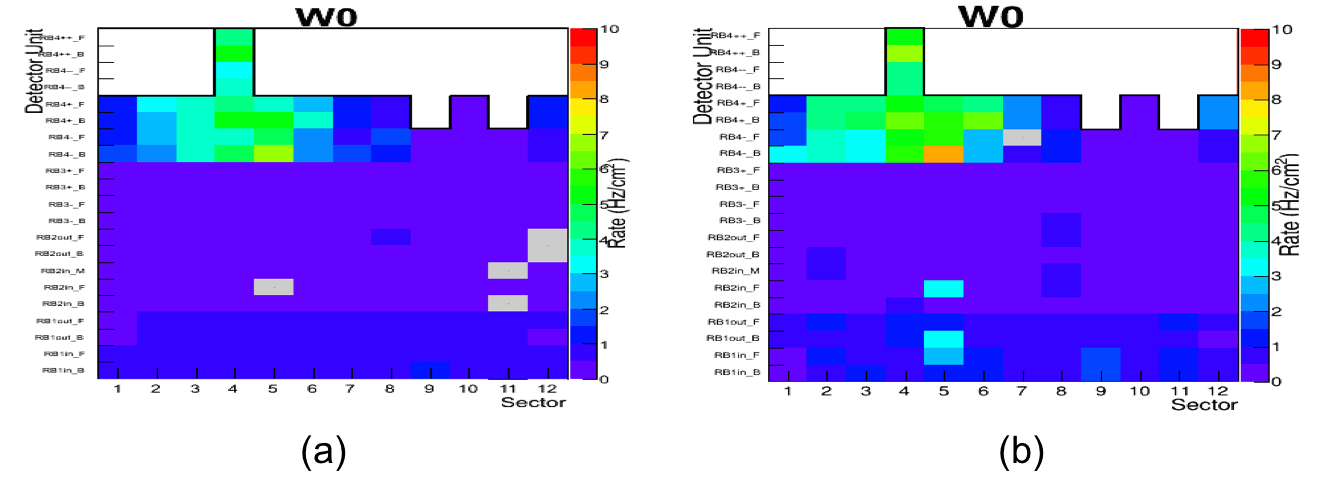}\\
\caption{The detector units hit rate (in Hz/cm\textsuperscript{2}) is shown for a run at average instantaneous  luminosity of 4.5*10\textsuperscript{33} cm\textsuperscript{-2}s\textsuperscript{-1}  for one of the barrel wheels in (a) at 8 TeV before 2013 and during 2015 in (b) at 13 TeV. Detector units switched off are shown in gray. Blue and violet colours  correspond to lower rates, while yellow, orange and red colours correspond to high background level.}
\label{fig:figure123}
\end{figure}
Low-momentum primary and secondary muons, low-energy gamma-rays, neutrons, and LHC beam-induced backgrounds could have an impact on performance of trigger and pattern recognition of muon tracks. 
\begin{figure}[H]
\centering
\hspace{-0.5cm}
\includegraphics[scale=0.6]{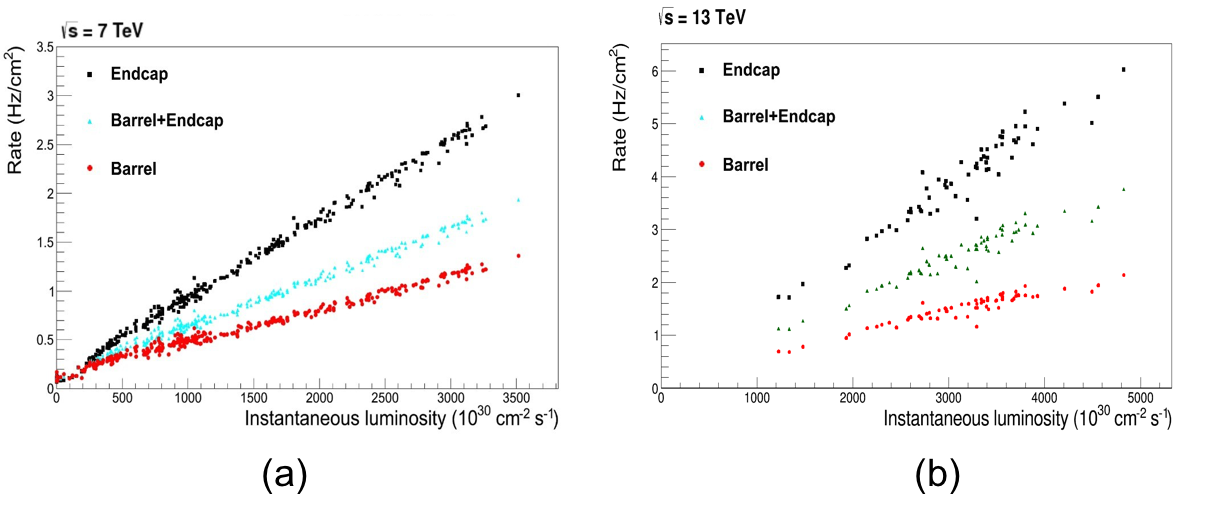}\\
\caption{The plots represent the average hit rate vs. instantaneous luminosity, with 2011 pp collisions at 7 TeV in (a) and 2015 pp collisions at 13 TeV in (b). The red dots represent the rate measured in barrel and the black represent the rate measured  in endcap. The green markers relate to the overall rate evaluated for the entire RPC system.}
\label{fig:figure1234}
\end{figure}
In addition, excessive radiation levels can also cause premature ageing of the detectors. Main contribution in measured RPC rate is coming from background. A plot is shown in figure~\ref{fig:figure123} as an example of 2015 at 13 TeV with its comparison to 2012 at 8 TeV. As shown in figure~\ref{fig:figure1234}, RPC rates increase approximately linearly with the luminosity of LHC. The linear behaviour can be used to extrapolate the rate for future upgrades. The rates increase for those chambers which are farther from the interaction point.%
\subsection{Efficiency vs. High Voltage}
A high voltage (HV) scan was performed every year: collision data was recorded at several HV settings during a series of runs to define the optimal operating voltage for each chamber. Details can be found in ~\cite{e, f} for a full explanation of the HV scan, dependence of efficiency on the HV, including the analysis and methodology. The dependence of the avalanche production on the environmental pressure P, temperature T and the applied HV can be summarised in an effective HV equation \eqref{eq:y:3}.%
\begin{equation}
HV_{eff} \left(P,T\right)\text{=}HV \left(P_0/P\right) \left(T/T_0\right)
\label{eq:y:3}
\end{equation}
Where HV$_{eff}$ is effective high voltage, HV is applied high voltage, and the reference temperature and pressure are T$_{0}$ = 293 K and P$_{0}$ = 965 mbar respectively.  
HV50 is defined as the high voltage at which a chamber reaches 50\% of the plateau efficiency.  
\begin{figure}[H]
\centering
\hspace{-0.5cm}
\includegraphics[scale=0.6]{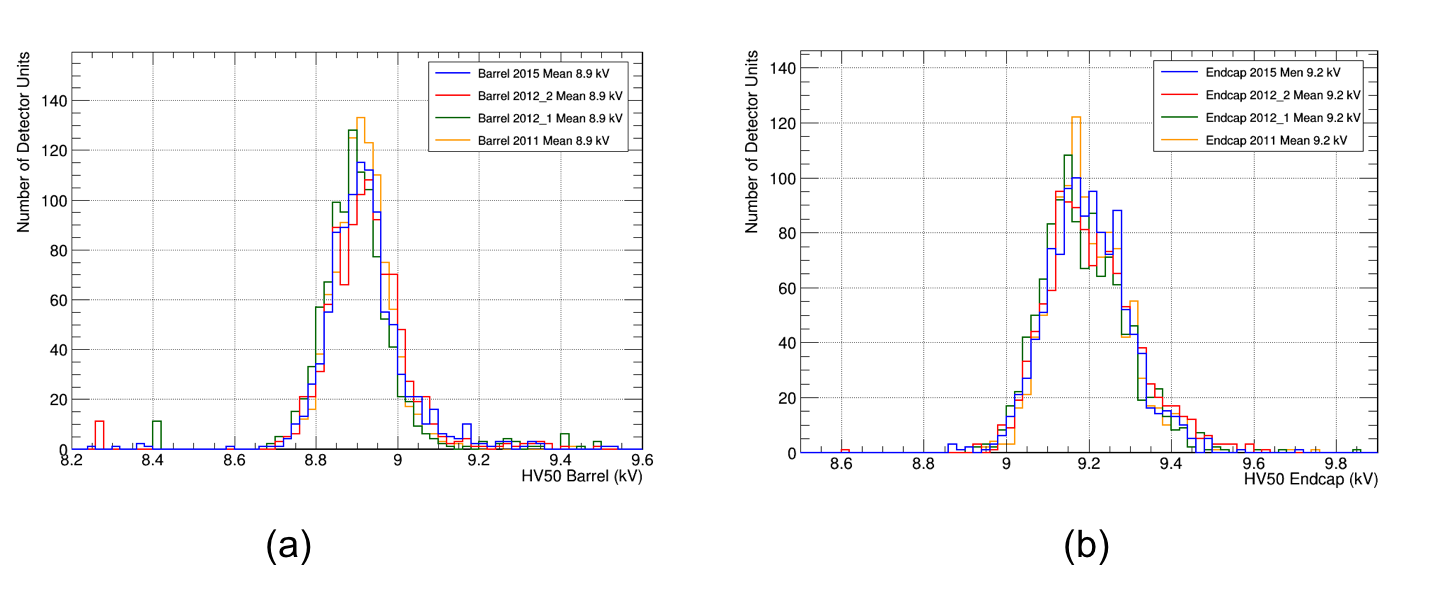}\\%\label{fig:subfigure1}}
    \caption{HV50 distributions for the barrel in (a) and for the endcap in (b) for 2011, 2012 and 2015.}\label{fig:animals111}
\end{figure}
 The width and the peak of distributions depend mostly on the construction specifications such as spacers sizes and operational conditions. The spacers are supports that create RPC gaps in the chambers. The distributions for 2011, 2012 and 2015 as shown in figure~\ref{fig:animals111} are very similar therefore no obvious ageing effect is observed. 
\subsection{Cluster Size}
\begin{figure}[H]
\centering
\hspace{-0.5cm}
\includegraphics[scale=0.6]{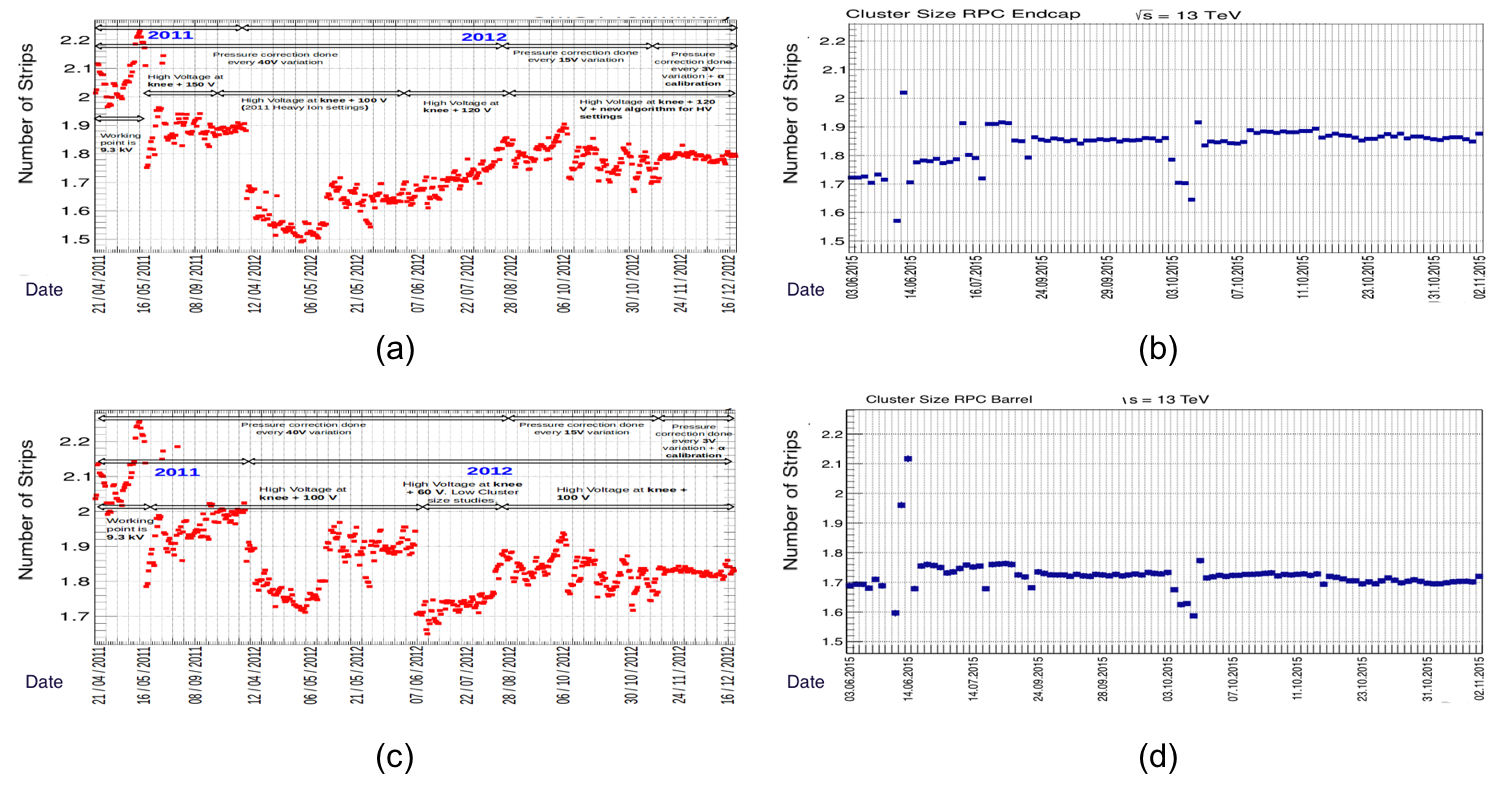}\\
\caption{The plots represent the history of the mean Cluster Size for the endcap and barrel for  2011 and 2012 physics data taking at 8 TeV in (a) and (c), and for 2015 at 13 TeV in (b) and (d).}
\label{fig:figure}
\end{figure}
Cluster size (CLS) is defined as the number of adjacent strips fired when an avalanche is produced in the RPC. RPC system has stable cluster size of about 1.8 strips over the years, which is in agreement with CMS TDR. CLS history in 2011 and in start of 2012 as shown in figure~\ref{fig:figure}, is affected by applied pressure corrections and several HV settings. During 2011 and the beginning of 2012 the applied HV to every RPC detector was corrected to compensate for pressure changes in the CMS cavern. The CLS at the end of 2012 was kept lower than 2011 to maintain a stable trigger rate. The fluctuation for 2015 in middle of June and beginning of October, are due to the performed HV and threshold scans.
\subsection{Efficiency}
\begin{figure}[H]
\centering
\hspace{-0.5cm}
\includegraphics[scale=0.6]{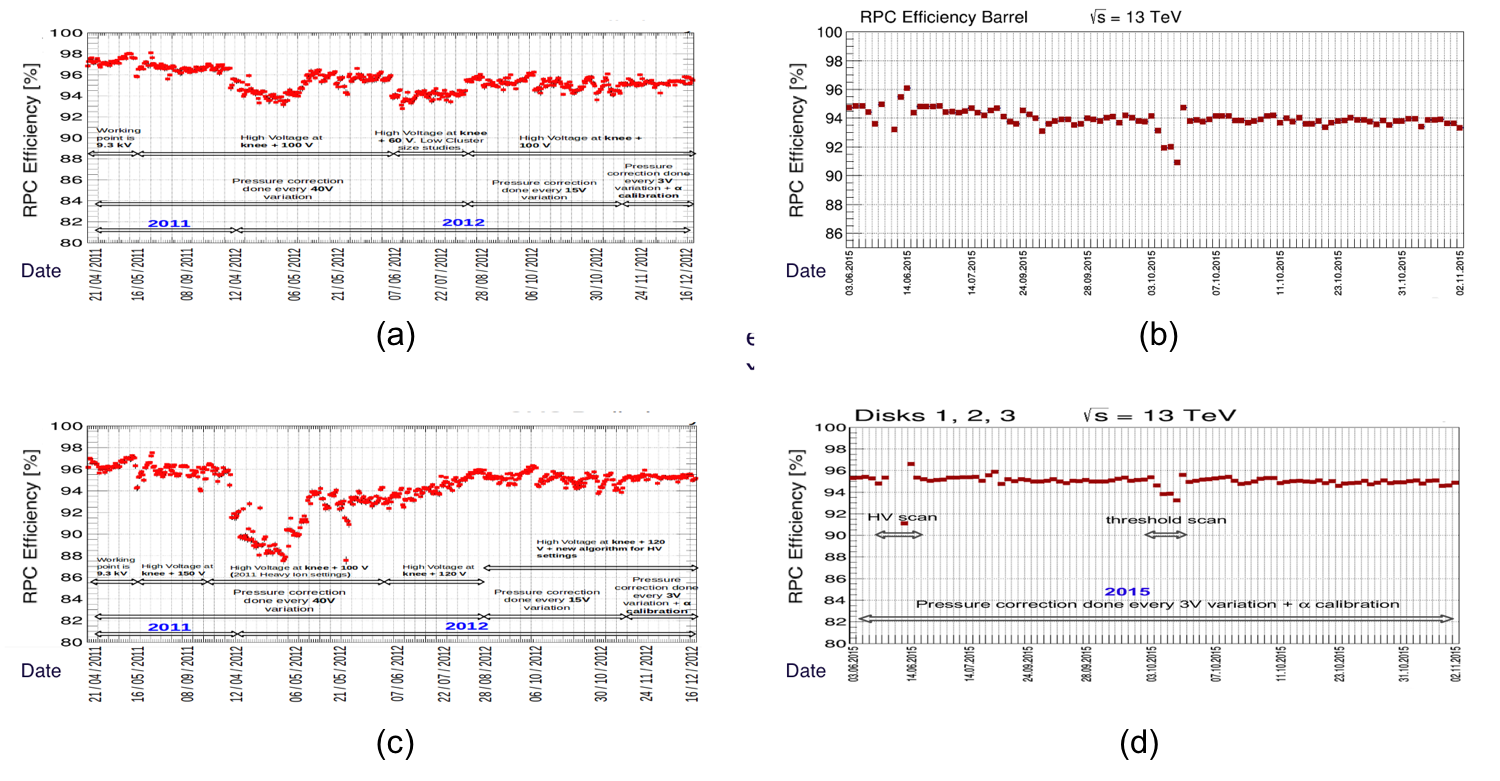}\\%\qquad
\caption{The plots represent the history of the overall RPC efficiency for the barrel and endcap for 2011 and 2012 physics data taking at 8 TeV in (a) and (c), and for 2015 at 13 TeV in (b) and (d).}
\label{fig:figure5}
\end{figure}

Segment extrapolation method ~\cite{g} is used to calculate the RPC efficiency. A DT/CSC segment of high quality, associated to a stand-alone muon track, is extrapolated to RPC strip plane. RPC efficiency depends on the atmospheric pressure in the cavern. In order to compensate this dependence, automatic corrections to the HV have been applied during the data taking. Efficiency is affected  by several HV settings and  applied pressure corrections, during 2011 and beginning of 2012 as shown in figure~\ref{fig:figure5}. The  fluctuation for 2015 in middle of June and beginning of October, are due to the performed HV and threshold scans.
Average RPC efficiency during 2015 at 13 TeV was $\approx$ 94$\%$ after 1 year of LHC running as detectors were operated at lower working points. During  2015 the RPC system was running with a very stable efficiency. 
\section{Conclusion}
CMS RPC system was operating very well during RUN-2 (2015). Performance is comparable with RUN-1 (2011-12) delivering good triggers and data for physics. After 1 year of LHC running with increasing instantaneous luminosity and 6 years from the end of RPC construction, the detector performance is within CMS specifications and stable with no degradation observed. From the measured background, no significant issues were found for running up to high luminosity scenarios. 
\acknowledgments We wish to congratulate our colleagues in the CERN accelerator departments for the excellent performance of the LHC machine. We thank the technical and administrative staff at CERN and all CMS institutes.

\end{document}